\documentclass[natbib, twocolumn]{svjour3}         
\smartqed  
\usepackage{graphicx}
\usepackage{mathptmx}      
\usepackage{aps-bibstyle}  
\newcommand{\alfven}{Alfv\'{e}n}
\newcommand{\alfvenic}{Alfv\'{e}nic}
\newcommand{\sdo}{{\em SDO}}
\newcommand{\hinode}{{\em Hinode}}

\newcommand{\RNum}[1]{\uppercase\expandafter{\romannumeral #1\relax}}
\newcommand\ion[2]{#1$\;${\RNum{#2}}}%

\newcommand{\pref}{\protect\ref}

\newcommand{\aap}{{Astron. Astrophys.}}

\newcommand{\apj}{{Astrophys. J.}}
\newcommand{\apjs}{{Astrophys. J. Supp.}}
\newcommand{\apjl}{{Astrophys. J. Lett.}}

\newcommand{\jgr}{{J. Geophys. Res.}}
\newcommand{\solphys}{{Solar Phys.}}
\newcommand{\ssr}{{Space Sci. Rev.}}
\newcommand{\pasj}{{Proc. Ast. Soc. Japan}}
\newcommand{\nat}{{Nature}}

\begin{document}

\title{The Evolving Magnetic Scales of the Outer Solar Atmosphere and Their Potential Impact on Heliospheric Turbulence}
\titlerunning{}

\author{Scott W. McIntosh  \and Christian Bethge \and James Threlfall \and Ineke De Moortel \and Robert J. Leamon \and Hui Tian}

\institute{Scott W. McIntosh \at
              High Altitude Observatory, 
              National Center for Atmospheric Research, \\
              P.O. Box 3000, Boulder, CO 80307 USA\\
              \email{mscott@ucar.edu}
              \and
              Christian Bethge \at
              High Altitude Observatory, 
              National Center for Atmospheric Research, \\
              P.O. Box 3000, Boulder, CO 80307 USA\\
              \email{bethge@ucar.edu}
              \and
              James Threlfall \at
              School of Mathematics and Statistics
              University of St Andrews,\\
              North Haugh, St Andrews, Fife, KY16 9SS UK\\
              \email{jamest@mcs.st-and.ac.uk}
              \and
              Ineke De Moortel \at
              School of Mathematics and Statistics
              University of St Andrews,\\
              North Haugh, St Andrews, Fife, KY16 9SS UK\\
              \email{ineke@mcs.st-and.ac.uk}
              \and
              Robert J. Leamon \at
              Department of Physics, 
              Montana State University,\\
              Bozeman, MT 59717 USA\\
              \email{leamon@mithra.physics.montana.edu}
              \and
              Hui Tian \at
              Harvard-Smithsonian Center for Astrophysics, 
              	60 Garden Street, \\
		Cambridge, MA 02138, USA\\
              \email{tian@head.cfa.harvard.edu}
              }

\date{Received: date / Accepted: date}

\maketitle

\begin{abstract}
The presence of turbulent phenomena in the outer solar atmosphere is a given. However, because we are reduced to remotely sensing the atmosphere of a star with instruments of limited spatial and/or spectral resolution, we can only infer the physical progression from macroscopic to microscopic phenomena. Even so, we know that many, if not all, of the turbulent phenomena that pervade interplanetary space have physical origins at the Sun and so in this brief article we consider some recent measurements which point to sustained potential source(s) of heliospheric turbulence in the magnetic and thermal domains. In particular, we look at the scales of magnetism that are imprinted on the outer solar atmosphere by the relentless magneto-convection of the solar interior and combine state-of-the-art observations from the Solar Dynamics Observatory (SDO) and the Coronal Multi-channel Polarimeter (CoMP) which are beginning to hint at the origins of the wave/plasma interplay prevalent closer to the Earth. While linking these disparate scales of observation and understanding of their connection is near to impossible, it is clear that the constant evolution of subsurface magnetism on a host of scales guides and governs the flow of mass and energy at the smallest scales. In the near future significant progress in this area will be made by linking observations from high resolution platforms like the Interface Region Imaging Spectrograph (IRIS) and Advanced Technology Solar Telescope (ATST) with full-disk synoptic observations such as those presented herein. 
\end{abstract}

\begin{figure*}
\begin{center}
\includegraphics[width=170mm]{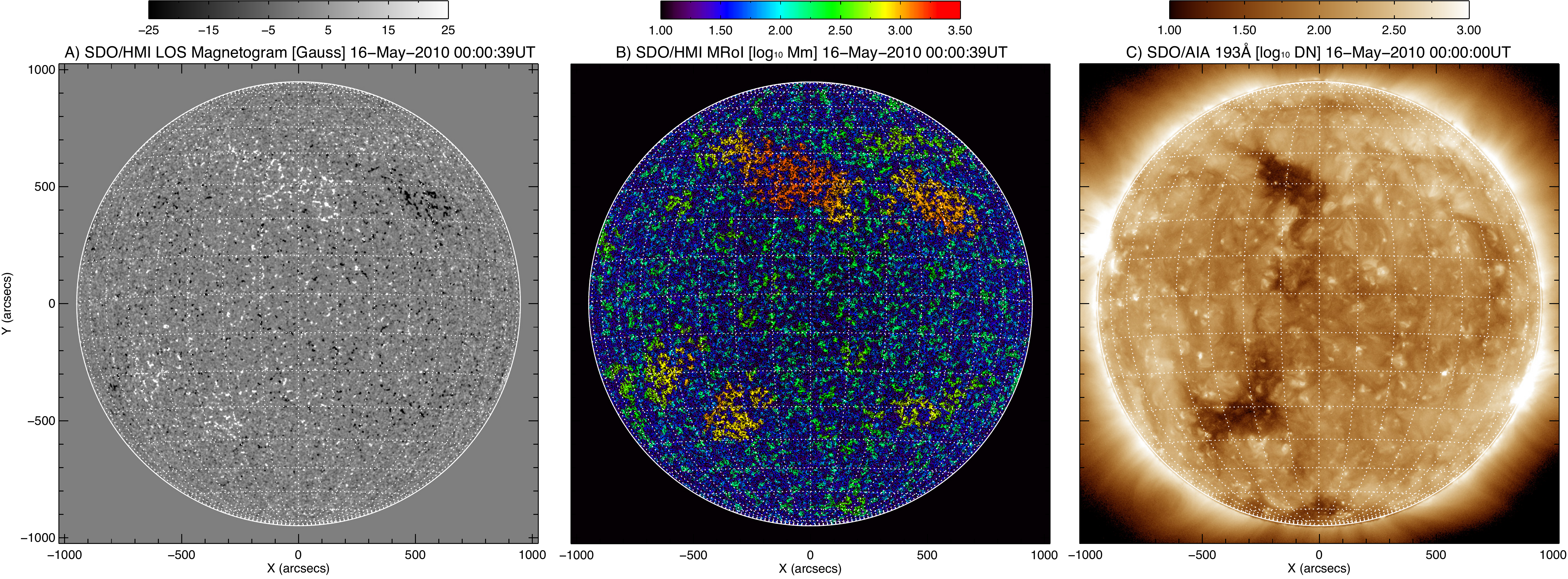}
\caption{From left to right, A) full-disk {\em SDO}/HMI line-of-sight magnetogram, B) the full-disk ÒMagnetic Range of InfluenceÓ (MRoI) image derived from the SDO/HMI line-of-sight magnetogram in panel A, C) the full-disk {\em SDO}/AIA image of the solar corona formed around 1.2MK in the 193\AA{} channel.}\label{f1} 
\end{center}
\end{figure*}

\section{Introduction}\label{intro}
The view of the outer solar atmosphere as series of stratified (spherical) layers is rapidly becoming outdated and has been replaced by a physical picture where it is organized by ``scales of magnetic connectivity'' dictated by the relentless turbulent evolution of the plasma beneath the solar photosphere (see the chapter by Valentin Martinez-Pillet). The so-called ``magnetic carpet'' \citep[e.g.,][]{1997ApJ...487..424S, 2003ApJ...597L.165S} and ``magneto-frictional" models \citep[e.g.,][]{1986ApJ...309..383Y, 2000ApJ...539..983V, 2000ApJ...544.1122M, 2012ApJ...757..147C} are (beautifully) illustrative constructs to demonstrate the progression of coupled magnetic evolution from the high plasma $\beta$\footnote{The ratio of the plasma pressure to the magnetic pressure, $\beta = 8\pi n k_B T / B^2$.} sub-photospheric driving and its impact on structuring of the low-$\beta$ outer atmosphere. 

In the atmosphere ``above'' the photosphere the evolution and organization appears on nested length scales. The smallest being granular cells with a diameter of a few Mm, a lifetime of a few minutes, and are easily discernible in the intensity and velocity patterns of the photosphere. Those granules are contained within supergranules which have a diameter of a few tens of Mm, lifetimes of about a day, and are most visible in the intensity and velocity patterns of the transition region \citep[e.g.,][]{1999Sci...283..810H,2007ApJ...654..650M}. There have been hints that a third scale of granulation exists, ``giant'' convective cells with a diameters ranging from 100 to 150Mm that are aligned with the Sun's rotation axis and lifetimes commensurate with the Sun's rotational period \citep[see, e.g.,][]{1998Natur.394..653B, 2008ApJ...673..557M}. The largest of these scales, while not having been unequivocally identified, has been linked to the emergence and evolution of active regions \citep[e.g.,][]{2011ApJ...741...11W}. 
We believe that this triumvirate of magneto-convective length-scales continuously evolve together to shape the passage of material and energy throughout the entire solar system \citep[e.g.,][]{2013arXiv1302.1081M}. 

In the not so distant future we will be able to bring observations of disparate connective and physical scales together to study how the relentlessly evolving magnetic scales merge, interact, and persistently drive the mass and energy flow into the heliosphere. Only then can much-coveted forecast models of solar activity and space weather be reliably built. Fortunately, observatories like the {\em Solar Dynamics Observatory} \citep[{\em SDO}; e.g.,][]{2011JGRA..116.4108S}, {\em Interface Region Imaging Spectrograph} ({\em IRIS}) and Advanced Technology Solar Telescope (ATST) will offer unparalleled insight into the connective physics of the solar atmosphere. However, we must start thinking now about how those scales connect, on what timescales, and what is the optimal set of observables required to assemble the complex picture consistently and with the least uncertainty. 

We are only beginning to understand the magneto-thermodynamic coupling of the Sun to the heliosphere. The first steps towards such a comprehensive picture (or model) involve a drive to understand the intricacies of the Sun's coupled atmosphere (interior and outer atmosphere), but the observations are complete, or mature enough, at this time to handle the range of spatio-temporal variability present. As a result of this technological limitation we are largely reduced to studying observational ``snapshots'' or isolated pieces of system evolution in the hope that fundamental processes and/or scales can be identified from them. This manuscript is neither a review nor research piece, presenting some interesting observational snapshots which merge analysis from different resolution instruments, illustrate the pertinent scales, and also perhaps indicate how, at the smallest scales, the seeds of heliospheric turbulence are sown.



\section{The ``Guiding'' Magnetic Scales of the Outer Solar Atmosphere}\label{mscale}
Before considering the injection of energy and mass into the heliosphere we take a look at the prevalent magnetic scales in the photosphere which shape the structure of the outer atmosphere and heliosphere. The ``Magnetic Range of Influence'' \citep[MRoI;][]{2006ApJ...644L..87M} is a construct used to study the magnetic length scales present in the outer solar atmosphere and was originally conceived to identify regions of extended (locally) unipolar magnetic field on the solar disk \-- ``coronal holes''.

The MRoI measures the (radial) distance from a given pixel we must travel outward such that the total magnetic flux in the area enclosed is zero. The MRoI can be considered as an extrapolation of the line-of-sight photospheric magnetogram without actually tracing the field lines. We follow the convention where Gauss' law holds, {\em i.e.}, that there are no magnetic monopoles, and all magnetic flux {\em must} close somewhere. The MRoI is constructed pixel-by-pixel and is a measure of magnetic balance, or the effective length scale over which we might expect the overlying corona to be connected. The full-disk {\em SDO}/HMI magnetogram of Fig.~\pref{f1}A is used to construct the MRoI map shown in Fig.~\pref{f1}B. By definition small values of MRoI indicate that the magnetic field there cancels (or closes) nearby, as is the typical case in the quiet Sun. Extended regions with larger values of MRoI ($\ge$500Mm) typically belong to active regions and coronal holes (see, e.g., with Fig.~\pref{f1}C) as originally identified in \citet{2006ApJ...644L..87M}.

\begin{figure}
\begin{center}
\includegraphics[width=85mm]{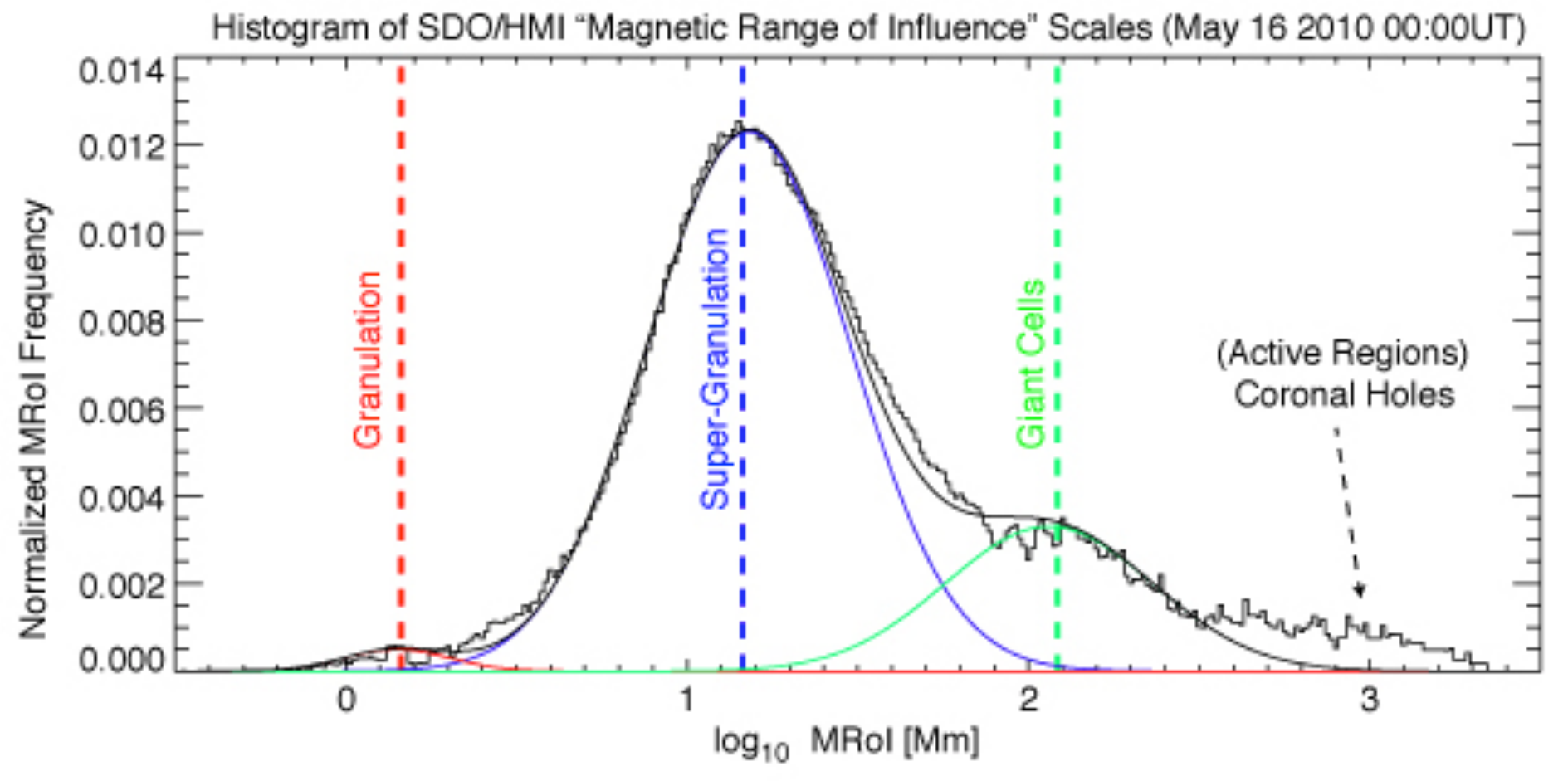}
\caption{Histogram of ÒMagnetic Range of InfluenceÓ (MRoI) derived from the full disk map of Fig.~\pref{f1}B. The histogram shows four possible scale ranges in the analysis: one near the resolution limit of {\em SDO}/HMI likely belonging to granulation (red dashed line), the scale of supergranulation $\sim$30Mm (blue dashed line), that of a 100-250Mm scale (green dashed line), and the longer length scales that we associate with coronal holes in this case.}\label{f2} 
\end{center}
\end{figure}

In an effort to extend this original work Fig.~\pref{f2} shows the histogram of MRoI values determined from Fig.~\pref{f1}B, and appears to show four overlapping components. The first component is small, just above the resolvable limit with {\em SDO}/HMI at a few Mm. The second (dominant) component of the MRoI distribution would appear to belong to magnetic fields canceling on the approximate scale of supergranules ($\sim$25Mm). As noted in the previous paragraph, visually comparing the {\em SDO}/HMI MRoI map with the {\em SDO}/AIA image of the corona, leads us to see that longest MRoI values of the distribution belong to coronal holes. Finally, there is the green component of the histogram. It is potentially very interesting, considerably smaller in frequency amplitude than supergranules, but contains MRoI values in the 100-200Mm range; consistent with that of giant cell convection which has been notoriously hard to definitively observe \citep[e.g.,][]{1998Natur.394..653B, 2005LRSP....2....1M}. This scale can be identified with the greenish colored clusters of MRoI pixels Fig.~\pref{f1}B. There is a possibility that these locations mark the vertices of a larger convective scale. 

We submit that the MRoI technique, when applied to full-disk magnetogram data, shows great promise as a diagnostic of the scales of atmospheric evolution. Further, if we can use MRoI analysis to routinely and robustly identify the apparent giant cell pattern in sequences of magnetograms then there is a chance that hypotheses of their role in the evolution of solar magnetism over timescales approaching those of the solar activity cycle can be directly tested \citep[e.g.,][]{1988Natur.333..748W, 2005LRSP....2....1M, 2008ApJ...673..557M}, but that is left as a topic for future investigation.

\begin{figure}
\begin{center}
\includegraphics[width=85mm]{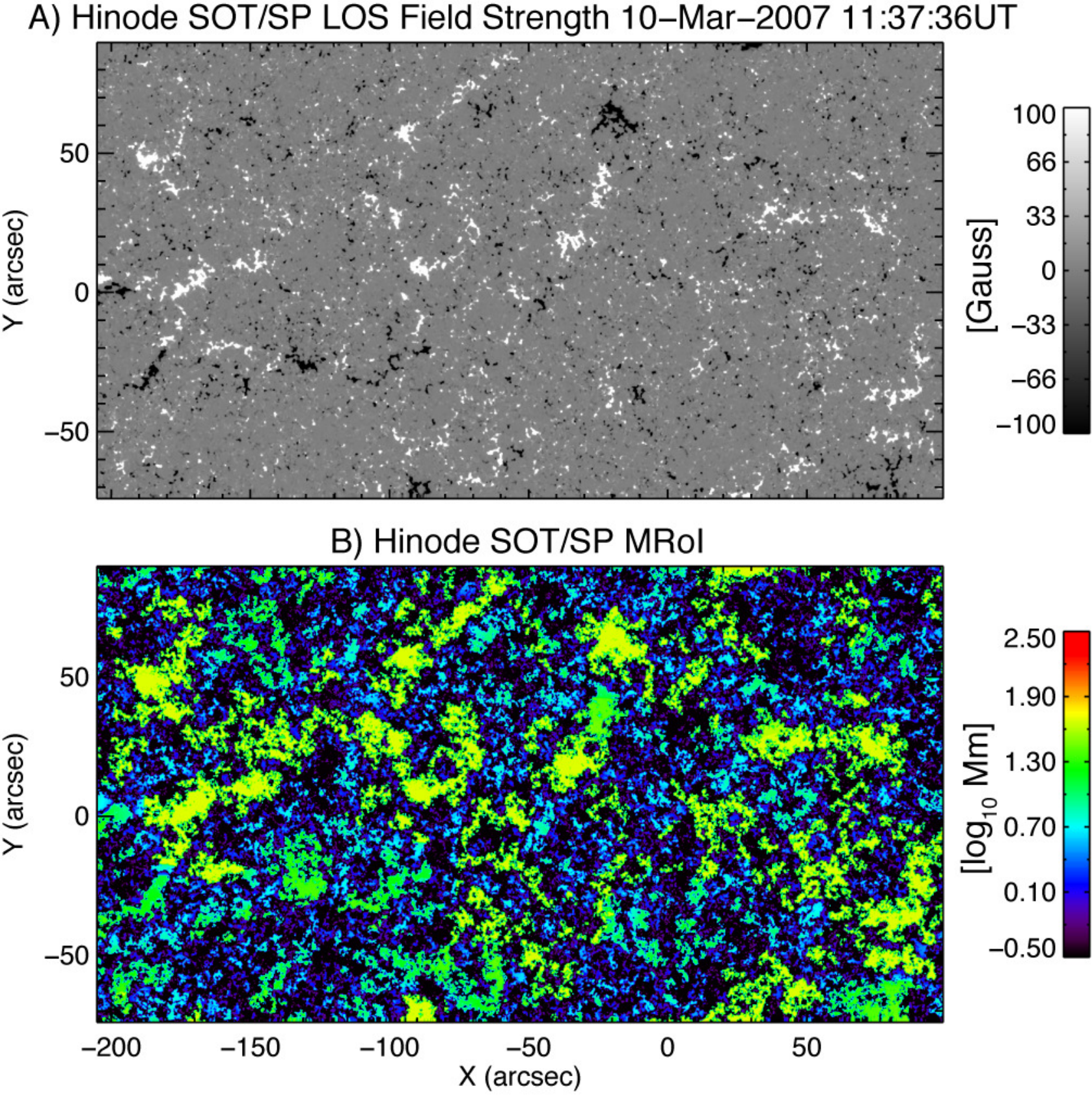}
\caption{The (inverted) line-of-sight magnetogram derived from {\em Hinode} SOT/SP (A) and resulting MRoI map (B) from observations taken on March 10 2007 at disk-center.}\label{f3} 
\end{center}
\end{figure}

Using line-of-sight magnetograms built from {\em Hinode} SOT Spectro-Polarimeter (SP) observations we are able to explore the smaller length scales at the other end of the spectrum - at or just below HMI's resolution limit. Figure~\pref{f3} shows an inverted\footnote{The SP maps are constructed by a forward of the observed Stokes vector at each pixel using the MERLIN algorithm and a Milne-Eddington atmosphere approximation \citep{2007MmSAI..78..148L}. The interested reader is pointed to \url{http://www.csac.hao.ucar.edu/} for further information.} SP line-of-sight magnetogram (A) and the resulting MRoI map (B) of a disk-center quiet sun (weak field) observation of March 2007 \citep[e.g.,][]{2008ApJ...672.1237L}.

\begin{figure}
\begin{center}
\includegraphics[width=85mm]{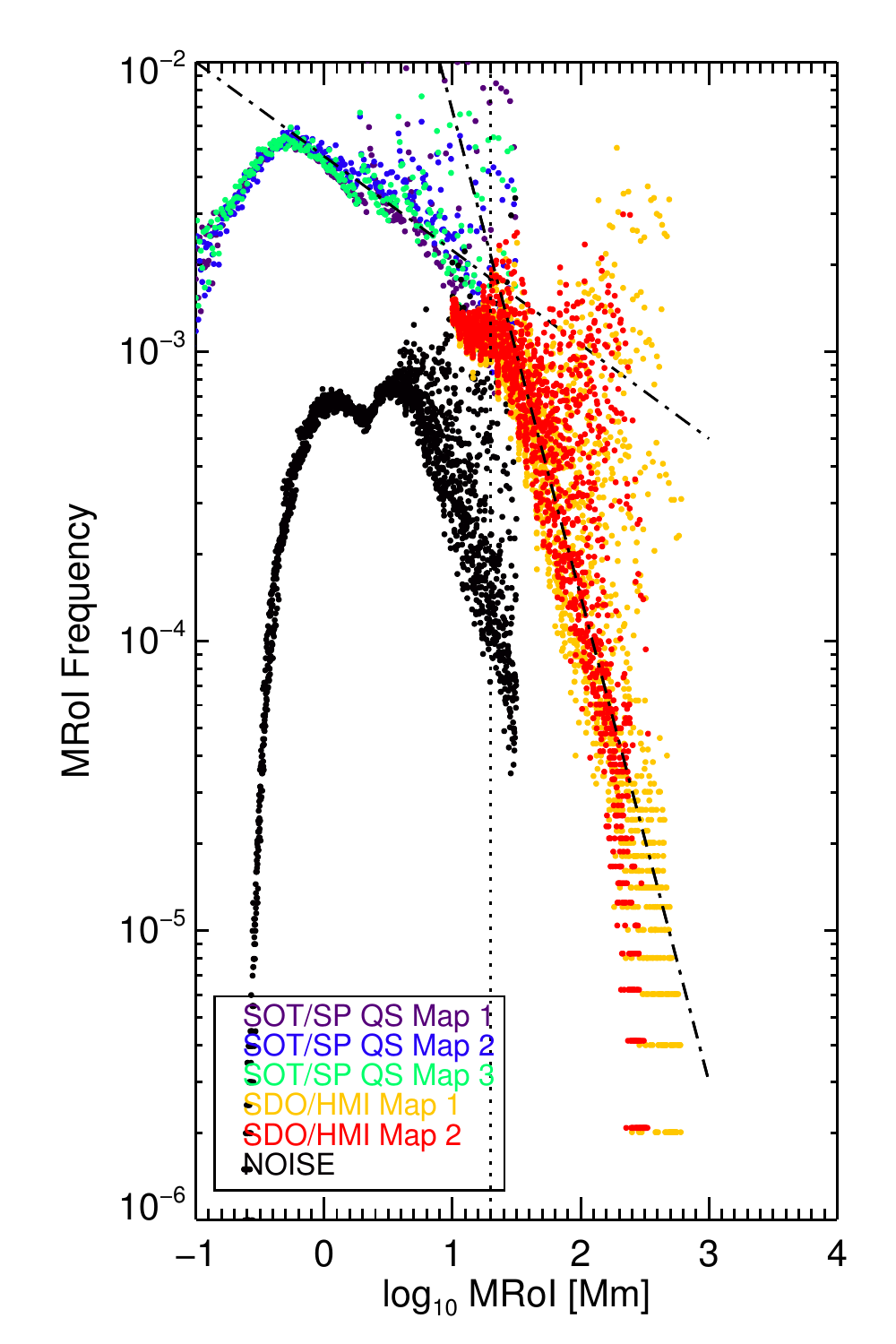}
\caption{Combining histograms of several MRoI maps built from {\em SDO}/HMI (orange, yellow) and {\em Hinode}/SP (purple, blue, and green) line-of-sight magnetograms. We see the continuity of the distributions across the scales and peaks in the combined distribution at the three prevalent scales of the quiescent magneto-convection. For reference we show an MRoI histogram constructed from a randomly populated magnetogram (black) using the same distribution of magnetic field strengths as shown in Fig.~\pref{f3}A.}\label{f4} 
\end{center}
\end{figure}

Combining histograms of SP and HMI MRoI maps in Fig.~\pref{f4} we see that these three magneto-convective scales appear to be ubiquitously present in the quiescent photosphere. The distributions have two different power-law distributions, with the spectral index between $0$ and $-1$ at shorter scales, steepening to between $-1$ and $-3$ for larger-scale magnetic structures, ignoring the bump around 100-250Mm. It has been shown in \citet{2011ApJ...730L...3M} that the longer scale index varies inversely over the course of the {\em SOHO} mission \citep{1995somi.book.....F}, shallowing at peak activity.

We propose that these three scales form the continuously evolving magnetic ``scaffold'' for the overlying coronal and heliospheric environment at scales of a few, tens of, and hundreds of megameters and for re-organization timescales of minutes, hours, and weeks to months. Coronal holes and active regions are the primary perturbations to these nested quiescent magnetic states. However, the continuing evolution of the base magnetism, at several spatial scales and at different rates leads to a plasma environment of constant interchange and reconnection which may ultimately make the extended corona more and more ordered. This continuous re-ordering of the nested scales to something quasi-potential at the longest length-scales is probably best thought in the context of ``interchange reconnection'' \citep{1996JGR...10115547F}. 

Unfortunately, coronal magnetic field measurements have not yet been refined to the level where the nascent coronal magnetic field, far less any turbulence present, can be directly observed, but instruments like the Coronal Multichannel Polarimeter \citep[CoMP;][and see below]{2008SoPh..247..411T} and the ATST/CryoNISRP will offer fantastic insight into the coupled scales of the resulting coronal magnetic field and its nested evolution.

\section{Loading Open Magnetic Regions of the Outer Solar Atmosphere}\label{oscale}

A great deal remains to be explored about the interplay of the magnetic length scales discussed above and their relevance to the variation of the Sun's radiative and particulate output. Combining instruments like {\em SDO}/HMI and the next generation of specto-polarimetric imagers (e.g., ATST/VTF) will permit a much more detailed look at the evolution of these scales and how their combination results in the persistent release of mass and energy into the extended outer atmosphere. The energy (and mass flux) that fills the Sun's outer atmosphere (and heliosphere) is driven by a host of physical mechanisms, often acting in unison \citep[e.g.,][]{2012RSPTA.370.3217P}. Those processes convert the sub-photospheric plasma motions on each of these spatial scales and likely scale in strength with the magnitude of the magnetic field at the vertices of the length scale itself \citep[e.g.,][]{1978ApJ...220..643R, 2004ApJ...615..512S}. So, with that picture and the analysis of the previous paragraphs in mind, we can think of the continuous loading of the outer atmospheric plasma as a result of the small-scale stress release in the plasma \citep[cf.][]{1988ApJ...330..474P} releasing energy gradually and relentlessly while the larger, more episodic, release of energy is the result of fractures in, or forced stressing to, the large scale field when a new magnetic system emerges into a pre-existing one \citep[e.g.,][]{2011JGRA..116.4108S,2013ApJ...768....8S}. In the following discussion we will look at processes which may combine to perpetually power the outer atmosphere on the smaller scales as result of this ``cascade''.

The energetics of the outer solar atmosphere are complicated: waves, flows, and magnetic reconnection are all likely to play a role in the process of delivering energy to the outer solar atmosphere, and eventually the heliosphere. Indeed, it is highly likely that these processes act sequentially, or simultaneously, in modifying the plasma {\em below the smallest scales discussed above} creating magnetic field  interactions which ultimately produce the photons or particles that we detect from millions of miles away. Therefore it is only natural that considerable debate takes place over the interpretation of those photons and particles, the following discussion is our interpretation and in full awareness that it may be neither unique or correct. 

To explore these small-scale processes we will make first make the assumption that the fundamental energetics of the entire quiescent solar atmosphere are essentially the same; the mechanisms observed in a coronal hole are virtually identical to those of the quiet Sun. So studying the base of coronal holes permits an investigation into the basal mass and energy injection of the quiet corona without the energetic complications that arise from a closed magnetic topology \citep[see e.g.,][and later]{2007ApJ...654..650M}. The (often subtle) observational differences between the two ``flavors'' of the quiescent solar atmosphere stem mostly from the fact that material trapped in the quiet corona has the ability to return to the lower solar atmosphere \-- a process that largely does not happen in coronal holes. The complete ``chromosphere-corona mass cycle'' of the quiet solar atmosphere modifies the spectral characteristics of the emission observed because both the emission components due to the (hot) upward mass flux {\em and} (cooling) return flow can be observed in a single spatio-temporal resolution element \citep[e.g.,][]{2012ApJ...749...60M}. Furthermore, it is expected that this mass cycle can affect the compositional differences that are measured in solar wind streams that originate from open and closed magnetic regions, but that is a topic far beyond this manuscript, and still at the frontier of our understanding.

The {\em SOHO} generation of spectroscopic studies have significantly improved our understanding of the relentless mass and energy transport at the base of the heliosphere \citep[see, e.g.,][and references therein]{1999A&A...346..285D,1999Sci...283..810H,2003A&A...399L...5X,2005Sci...308..519T, 2006ApJS..165..386D,2006ApJ...644L..87M}. Recently, a picture of energy release and initial mass supply to the corona and solar wind originating in the lower atmosphere has developed following the observation of a finely scaled, high velocity component of the spicule family \citep{2007PASJ...59S.655D}. These ``Type-II'' spicules have been associated with a weak but ubiquitous signature of chromospheric mass supply to the outer solar atmosphere rooted in unipolar magnetic field regions \citep[through the study of transition region and coronal spectroscopy, e.g.,][]{2009ApJ...701L...1D,2009ApJ...707..524M,2010A&A...521A..51P,2010ApJ...722.1013D, 2011ApJ...727L..37T} and broadband coronal imaging sequences \citep[e.g.,][]{2009ApJ...706L..80M}. 

\begin{figure}
\begin{center}
\includegraphics[width=65mm]{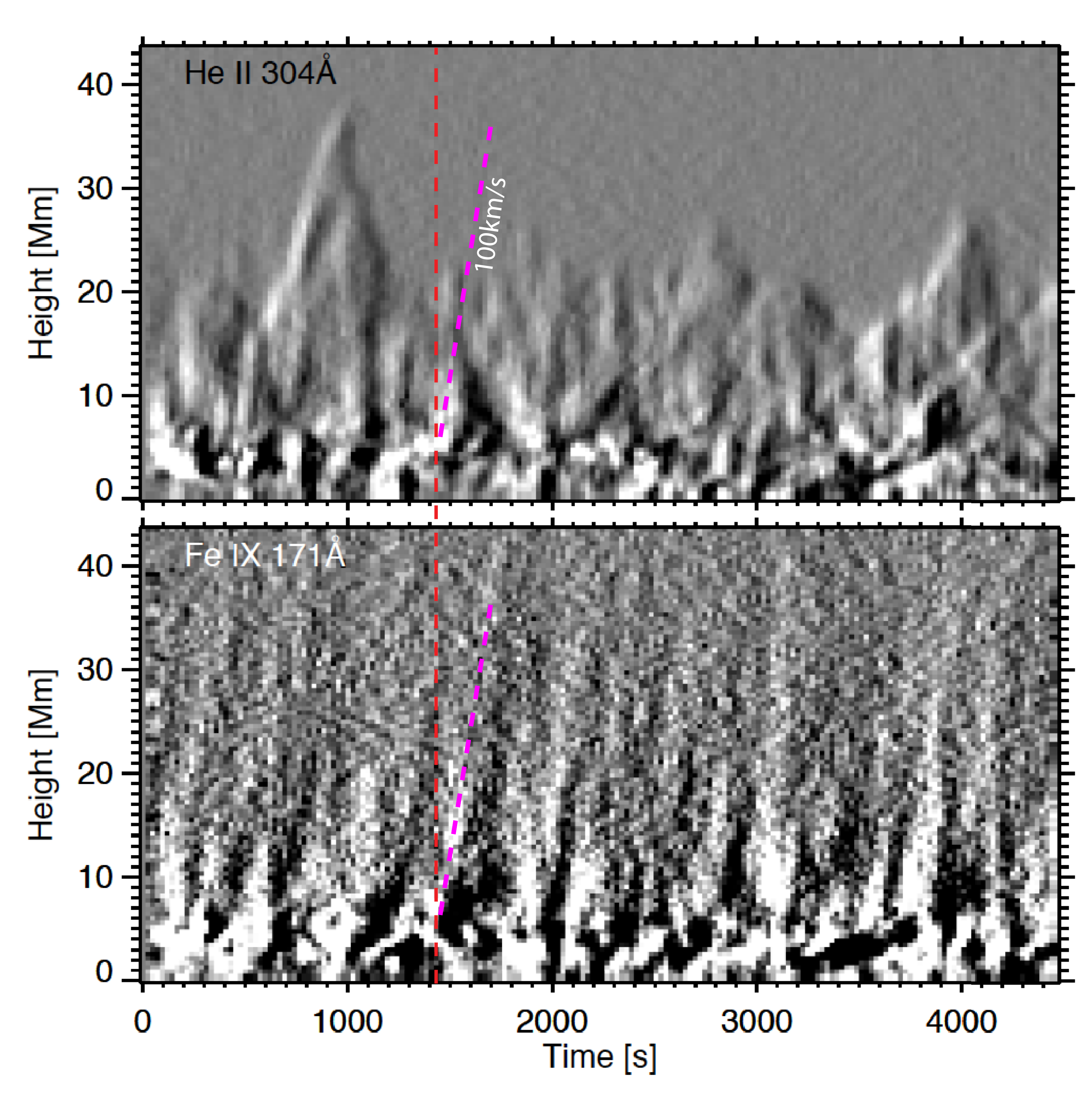}
\caption{Adapted from \citet{2011Sci...331...55D}. The temporal and thermal evolution of off-limb spicules in a coronal as observed in the \sdo/AIA \ion{He}{2} 304\AA{} (top) and \ion{Fe}{9} 171\AA{} (bottom) passbands, summing three images in time to boost the visibility of the faint off-limb signals, with a resulting cadence of 24s. Calculating space-time plots for a ``cut'' perpendicular to the limb, i.e., parallel to the prevailing spicules and the resulting space-time plots of the running difference (over 3 timesteps, i.e., 72s) show many clear parabolic (up and down) paths in \ion{He}{2}  304\AA{} that are often associated with upward propagating disturbances in \ion{Fe}{9} 171\AA{}  (PCDs)---the red vertical dashed line highlights one spicule/PCD correspondence. The \ion{He}{2} 304\AA{} spicules have lifetimes of order several minutes. The apparent propagation speed of the PCD and initial phase of the spicules are of order 100 km/s (the pink dashed line is shown as a reference).}\label{f5} 
\end{center}
\end{figure}

The detailed correspondence of dynamic chromospheric and coronal plasma heating has recently been confirmed by joint observations of {\em Hinode} \citep{2007SoPh..243....3K} and \sdo{} that are presented by \citet{2011Sci...331...55D}. In that paper discrete high-velocity heating jets triggered in the lower-atmosphere have been uniquely traced through the chromosphere, transition region, and into the outer atmosphere (as ``Propagating Coronal Disturbances'' or PCDs) for the first time (see Fig.~\pref{f5}). PCDs are detected in coronal holes using space-time (X-T) plots to monitor the evolution of the plasma in the first few Mm perpendicular to the limb in the \ion{He}{2} 304\AA{} (top) and \ion{Fe}{9} 171\AA{} (bottom). We see that there are many instances where the initial phase of the transition region spicule (see the red dashed vertical line as an example) is shared with the launch of the PCD.  Indeed, both features have the same initial speed (the pink dashed line indicates an apparent speed of 100km/s).

\begin{figure}
\begin{center}
\includegraphics[width=85mm]{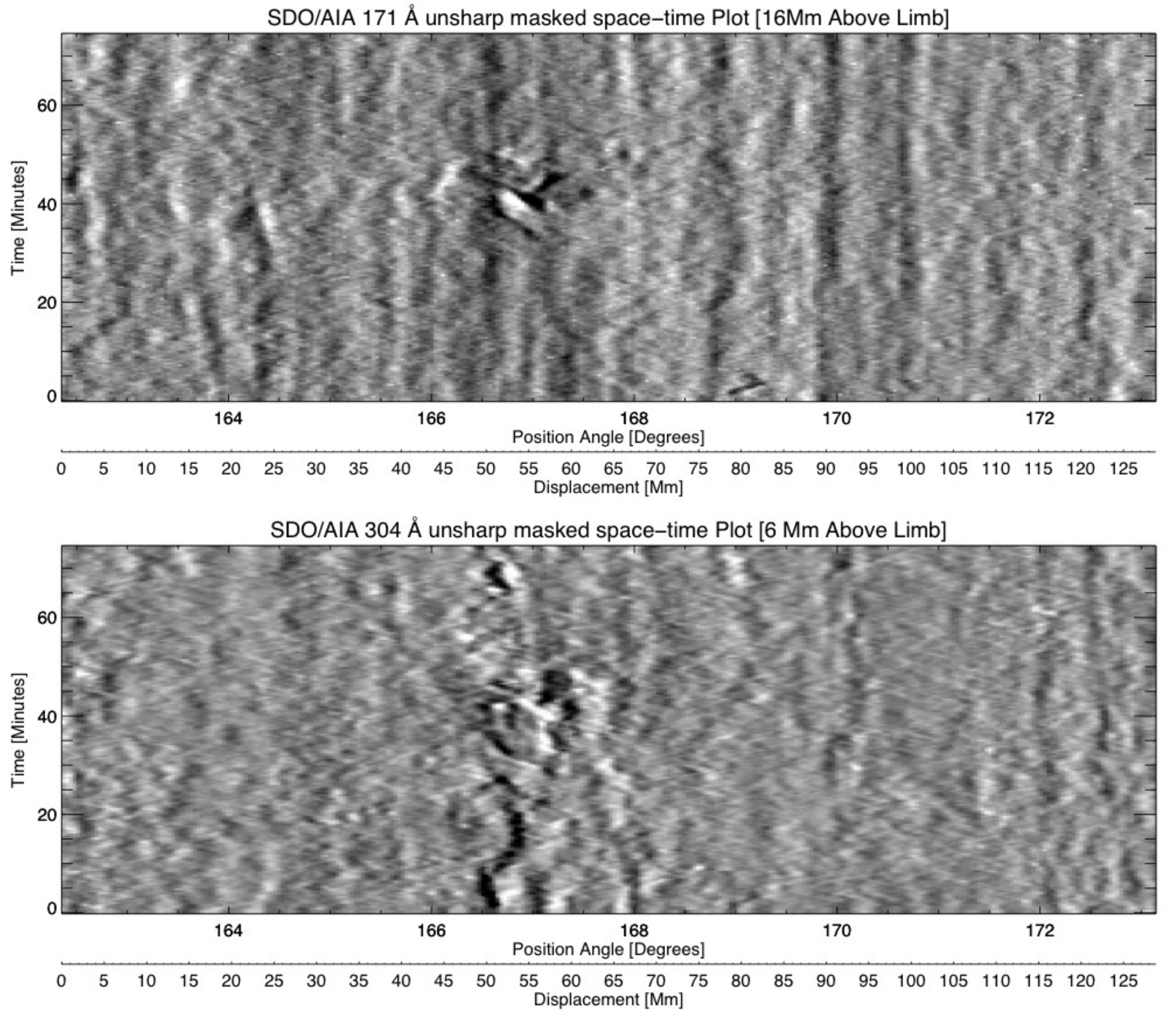}
\caption{Adapted from the supporting online material of \citet{2011Natur.475..477M}. The transverse motion of transition region spicules (bottom) and associated PCDs (top) derived from in space-time plots taken parallel to the solar limb of a polar coronal hole. The \alfvenic{} motions are rarely complete sinusoids because of the short spicule/PCD lifetimes. We also see wave motions that are horizontally separated by as much as 5~Mm that have in-phase transverse oscillations. This provides support for our assertion that the waves are volume filling.}\label{f6} 
\end{center}
\end{figure}

Another recent observational development relevant to the basal energetics of the fast solar wind was the identification of a sufficient \alfvenic{}\footnote{The interested reader is pointed to the supporting material of \citet{2011Natur.475..477M,2011Natur.475..463C} where a lengthy discussion defines their use of this term to describe the observed motion, relative to the theoretical terms ``kink'' or ``\alfven'' which require a more precise description of the plasma and its magnetic field than can be determined observationally.} wave energy flux to propel the fast wind. inferred that an abundant \alfvenic{} wave flux was present in the chromosphere, visible in the motion of {\em all} chromospheric spicules, but particularly clear to observe in the polar coronal holes that are dominated by Type-II spicules. Due to the lack of coronal observations with commensurate spatio-temporal resolution with \hinode{}, it was unclear at that time whether or not the observed wave flux was reflected at the chromosphere-coronal boundary. A subsequent observational study by \citet{2011Natur.475..477M} used \sdo{}/AIA observations to infer that the wave amplitude and phase speed throughout the solar atmosphere through the motions of transition region spicules and PCDs at coronal temperatures (see Fig.~\pref{f6}). The observed 100-200Wm$^{-2}$ of wave energy in coronal holes resided in \alfvenic{} motions of 300-500s period, 25km/s amplitude and 1Mm/s phase speed --- a flux of waves streaming outward on structures that weakly emit at temperatures of order 1MK\footnote{The emission in which the PCDs are most clearly observed is formed (in equilibrium) at $\sim$1MK. It is unclear given the rapid nature of the heating readily visible at the base of these structures (see below) whether or not the plasma is in ionization equilibrium, or not, during the initial heating phase. It is likely that some of the material is in excess of 1MK upon breaching the Sun's atmosphere.}. This wave flux is sufficient to drive the fast solar wind \citep[e.g.,][]{1995JGR...10021577H,2012ApJ...761..138M}.

As we have seen above, the AIA \ion{He}{2} 304\AA{} channel at the solar limb shows a transition region that is dominated by spicular jets that shoot rapidly upwards, and often those jets reach heights of 20,000km above the solar limb (Fig.~\pref{f5}). Observations of the same region in the \ion{Fe}{9} 171\AA{} channel reveal associated PCDs that propagate outward at high speeds ($\sim$100 km/s). When studying space-time plots parallel to the solar limb it is clear that these transition region and coronal features undergo significant \alfvenic{} transverse motion with displacements varying sinusoidally in time \citep[see, e.g., Fig.~1 of][]{2011Natur.475..477M}.

Indeed, SDO/AIA image sequences of polar coronal holes show an outer atmosphere that is replete with \alfvenic{} motion. The waves are traced by structures that do not have long lifetimes (of order 50-500s) compared to the wave periods ($\sim$5~minutes), and are difficult to detect because of the enormous LOS superposition above the solar limb. These factors contribute to the fact that very few complete swings of the spicule (or PCD) are observed and we are left with the ``criss-cross'' pattern of temporal evolution at a specific height above the limb (Fig.~\pref{f6}). \citet{2011Natur.475..477M} followed the analysis of \citet{2007Sci...318.1574D} and used Monte Carlo simulations to study the patterns produced by the propagation of the of the transverse motion. They found that the coronal hole waves had periods in the range of 150-550s, and amplitudes of order 25km/s in emission characteristic of coronal temperatures, i.e. clearly some portion of the \alfvenic{} energy had made it ``into the corona'' without being reflected. Using cross-correlation techniques of parallel space-time plots at different heights above the polar limb to determine the phase speed of the \alfvenic{} motions (reaching 1~Mm/s at an altitude of 50Mm) \citet{2011Natur.475..477M} identified that the volume filling waves carried somewhere between 100 and 200 Wm$^{-2}$ at the lower boundary of the fast solar wind.

Putting these observational findings together, we have the distinct possibility that (at least) two mechanisms at work in the magnetic elements that comprise the tributaries of the fast solar wind. First is the (quasi-periodic) plasma heating and material injection to temperatures of order 1MK visible in the spicules and PCDs, which is augmented by a secondary significant flux of long period \alfvenic{} motions that are needed to accelerate that mass away from the Sun. We suggest that somehow, this two-stage ``engine'' could be responsible for loading the open quiescent atmosphere. Interestingly, recent papers have also characterized the considerable torsional motions that are also present in Type-II spicules \citep[][]{2012ApJ...752L..12D, 2013ApJ...764..164S}, and may also transport a significant flux of twist and mechanical energy into the outer atmosphere \-- this is an exciting new avenue of exploration, {\em only} open to studies of the chromosphere made by high-resolution spectral imagers such as DST/IBIS, SST/CRISP and soon ATST/VBF.

It is our view that the fine details of the fast wind initiation and acceleration mechanisms are within reach, as the observations and simulations are moving forward at pace. The findings noted above form the simplest case, and the solution of {\em that\/} simple puzzle could yield invaluable insight into the complex evolution of the closed magnetic regions. The ``complication'' added in the case of the closed corona (and the slow solar wind that originates there), requires that we {\em must} consider the circulation of material between the chromosphere and corona and other factors {\em before} we consider the transition of that material onto an open field line and outward into interplanetary space \citep[e.g.,][]{1996JGR...10115547F}.

\begin{figure*}[!t]
\begin{center}
\includegraphics[width=160mm]{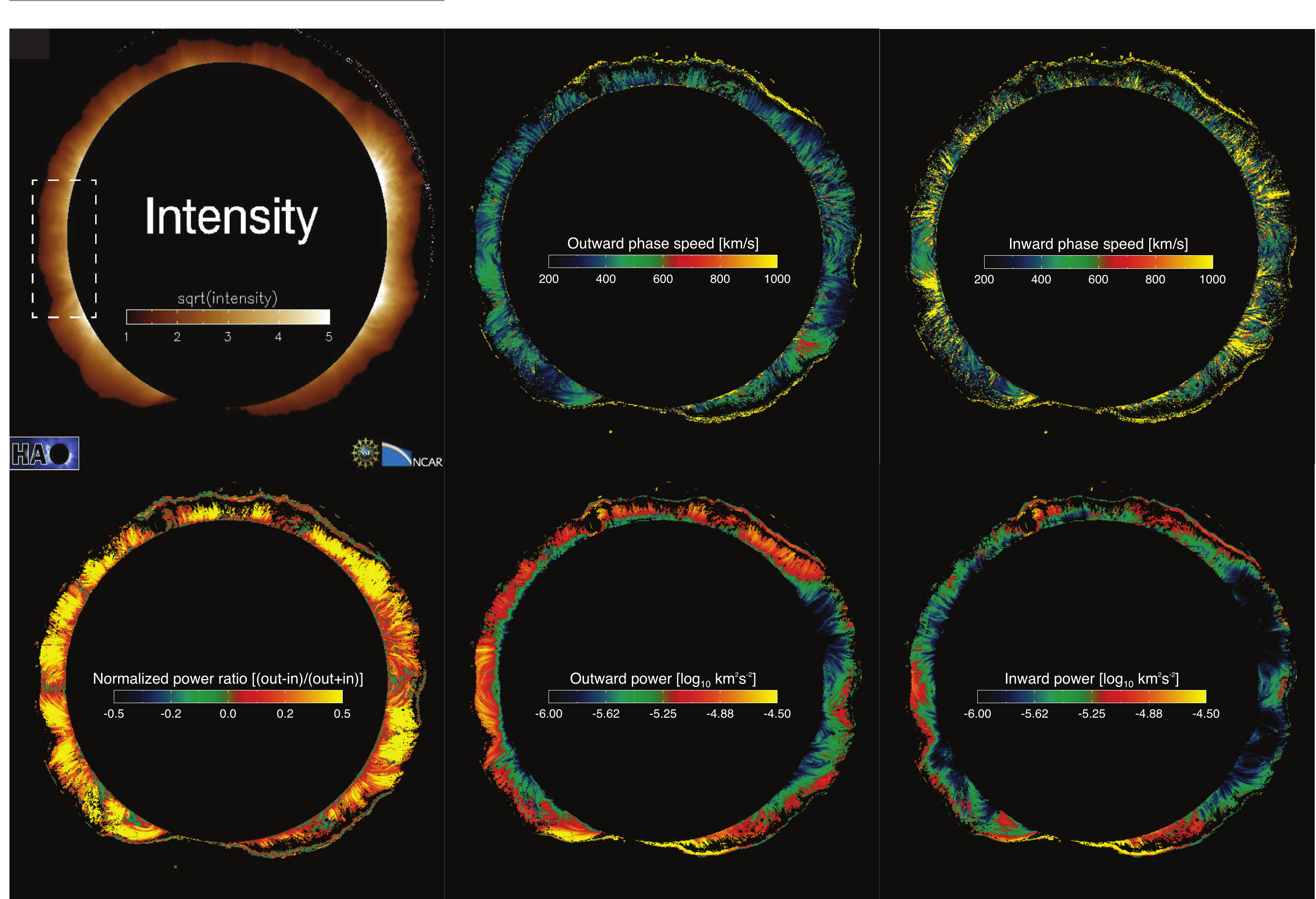}
\caption{April 10, 2011 diagnostics from the Coronal Multi-channel Polarimeter (CoMP). From top to bottom, left to right, we have the Fe~XIII 10747\AA{} line center intensity, the measured outward and inward wave phase speeds, the normalized wave power, the individual outward and inward wave power. The rectangular box on the top left panel indicates the location of the detailed plots shown in Fig.~\pref{f8} which isolate the waves on the large trans-equatorial coronal loops.}\label{f7} 
\end{center}
\end{figure*}

\section{Loading Closed Magnetic Regions of the Outer Solar Atmosphere}

If our earlier assumption that the prevalent mechanisms driving the mass and energy flow into coronal holes are the basic building blocks of the quiescent outer atmosphere (if not the entire outer atmosphere) then closed magnetic regions {\em should} have wave and mass loading taking place at {\em both} ends of the structure. Therefore, one may naturally expect that some form of amplification/interaction will take place more rapidly and with increased complication in the mid-space of the structures. In the following paragraphs we will see if this natural extension to closed regions is viable, and what the implications are for future observations?

Following the \hinode{} SOT observations which discovered the high velocity Type-II spicules an investigation started to identity if the rapid fading and upward motion of these very dynamic events was a marker of intense plasma heating in the lower atmosphere. \citet{2009ApJ...701L...1D} inferred a connection between the apparent motion of Type-II spicules and a weakly emitting component of emission observed far in the blue wing of coronal emission lines formed at a broad range of temperatures, reaching at least 2MK \citep[as observed by \hinode{}/EIS,][]{2007SoPh..243...19C}\footnote{This artificial upper bound is driven by limited observation of relatively ``clean'' spectroscopic emission lines in the wavelength range studied by EIS.}. This subtle signature (at 5-15\% of the background emission) was only visible in the highest signal-to-noise spectroscopic datasets from \hinode{}, but it appeared to have a very strong effect on the observed ``non-thermal'' broadening of the line profiles over magnetized regions \citep[e.g.,][]{2008ApJ...678L..67H}. \citet{2009ApJ...701L...1D} developed a diagnostic measure to investigate the excessive broadening of the coronal emission lines by differencing the amount of emission in the blue and red wings of a line profile. The difference in the two was interpreted as a net imbalance of upward (blue) or downward (red) emission at that velocity (relative to the measured center of the line profile). A residual high speed component of the emission was observed to coincide with the magnetized locations.

The high S/N, low scattered light of the {\em SDO}/AIA telescopes permitted the combination of these line profile asymmetries with the detailed imaging of the high velocity upwards propagating events (cf. PCDs). These imaging investigations follow from the analysis of \citet{2007Sci...318.1585S} who identified episodic, high speed flows on coronal loop structures in lower spatial resolution \hinode{} XRT observations that were rooted in the underlying strong magnetic field regions \citep[also see][for details of the correspondence between these events and the spectroscopic observations of line profile asymmetries]{2009ApJ...707..524M, 2011Sci...331...55D, 2011ApJ...738...18T,2012SoPh..279..427K}. \citet{2011Sci...331...55D} demonstrated that Type-II spicules, PCDs, and blue-wing asymmetries were co-spatial and likely markers of energy release and resulting mass transport in the closed corona of sufficient magnitude and frequency to satisfy coronal mass requirements up to 2MK\footnote{Unfortunately, the S/N in the ``hotter'' AIA passbands is significantly lower, and so it is unclear what the maximum temperature of the material in the PCD is.}. Even with the significantly better observing conditions of {\em SDO}/AIA, establishing a one-to-one connection between spicule and PCD is difficult.

The PCDs exhibit complex behavior at the base of the entire system: longitudinal motions of the order of 100km/s, repeat times of a few to fifteen minutes, transverse wave amplitudes of the order 25km/s which have phase speeds of several hundred km/s and periods in the 3-5 minute range \citep{2011Natur.475..477M}. Naturally, these properties can vary dramatically between the quiescent and active portions of the closed atmosphere, but a systematic survey remains to be done. Unfortunately, however, the transverse (\alfvenic) motions of the PCDs are {\em very} difficult to study even just a few Mm above the photosphere and so we must consider a different way to study the extended properties of these potentially critical combination of material and wave motions (in the context of atmospheric and heliospheric energetics).

The Coronal Multi-channel Polarimeter, or CoMP, was designed to directly measure coronal magnetism through the observation of polarized radiation from coronal emission lines. Fortunately, due to its design as an imaging spectropolarimeter, CoMP was able to perform rapid spectral scans of coronal emission lines in a two-dimensional image with a field of view of 2.5 solar radii in only a few seconds. These observations led to the first observations of \alfvenic{} waves running though the off-limb corona at speeds approaching 1Mm/s \citep{Tomczyk2007}. Subsequent investigations with the same dataset showed that a much richer \alfvenic{} environment exists in the corona: the waves observed could be differentiated into their inward and outward propagating components by developing time-distance coronal seismology and that there was a systematic bias (3:1) between the power in the outward and inward wave power \citep{2009ApJ...697.1384T}.

Figure~\pref{f7} provides an array of CoMP diagnostics measured using the diagnostics of the Fe~XIII 10747\AA{} on April 10, 2011. Identifying the waves and tracking their motion through the corona allows us to assess the wave properties of the corona. While the diagnostics are limited by low signal-to-noise in open regions of the corona for Fe~XIII we see that the wave phase speeds are in the 400-500km/s range in the closed field regions and that the outward wave power is considerably higher than the inward component in general, confirming the primary analysis of \citep{2009ApJ...697.1384T}. {\em Clearly, not all of the appreciable wave power that leaves the lower solar atmosphere is able to propagate across the coronal loop systems and maintain its original form.} It is not clear if the waves are dissipated in situ, or if they are converted into another ``mode'' which CoMP can no longer trace \citep{2012A&A...539A..37P,2013A&A...551A..39H,2013A&A...551A..40P}. 

We note that the temporal variations in line intensity and line width are very small \citep[$<$1\% and $<$1km/s; cf.][]{Tomczyk2007} and are difficult to interpret due to the three-point measurement needed when CoMP is taking high-cadence wave measurements. The wave diagnostics shown in the figure are being computed on a daily basis to enable a synoptic survey of the gross coronal (\alfvenic) energetics.

\begin{figure}
\begin{center}
\includegraphics[width=85mm]{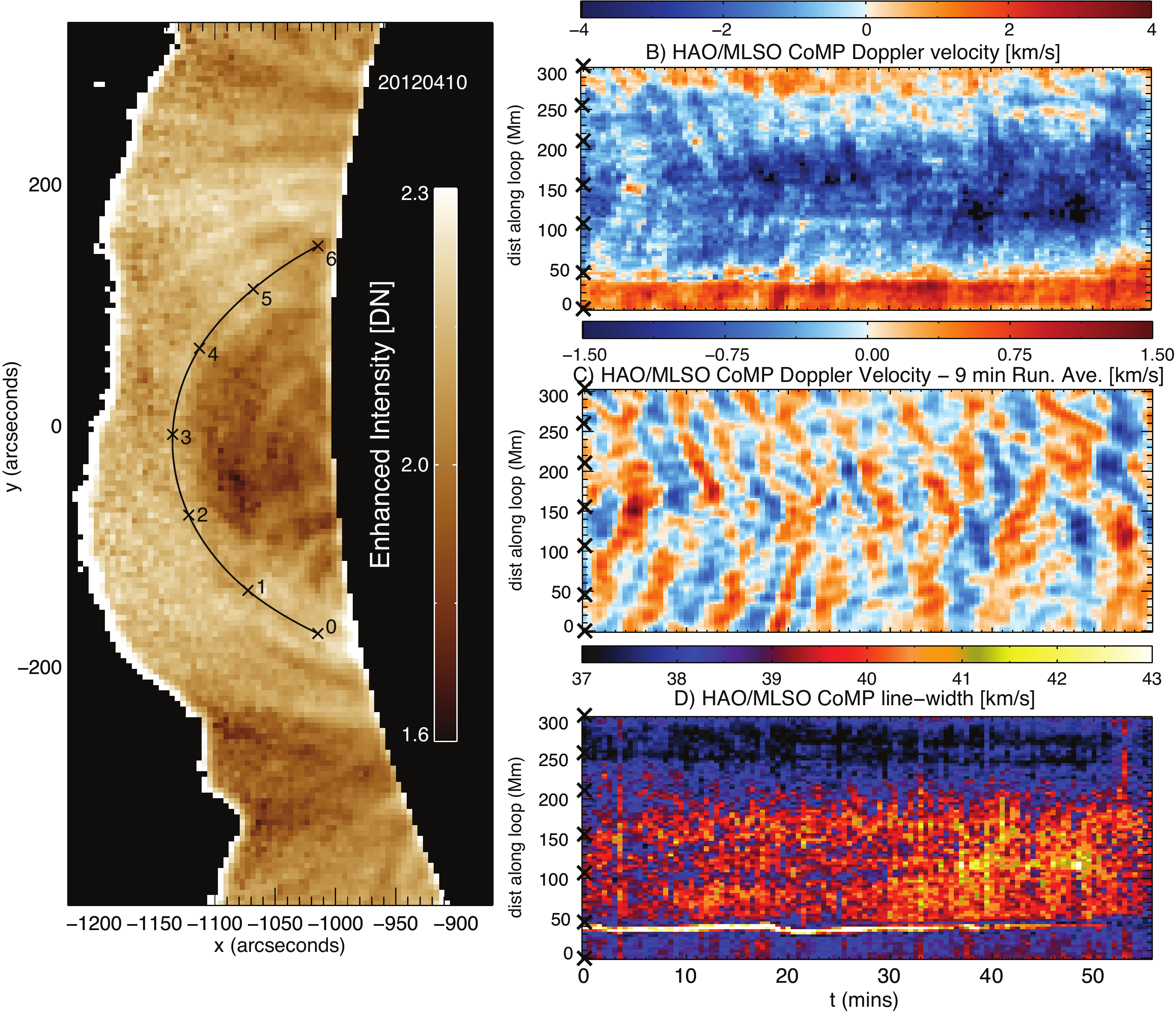}
\caption{Space-time plots showing \alfvenic{} waves on a trans-equatorial coronal loop system. The left panel provides a closer look at the loop system (indicated as the rectangular region in Fig.~\pref{f7}). The right column of the figure shows the Fe~XIII 10747\AA{} Doppler velocity, the trend-removed Doppler velocity, and the timeseries of line width variations along the loop.}\label{f8} 
\end{center}
\end{figure}

Figure~\pref{f8} provides more detail of the trans-equatorial loop system on the East limb (shown in the rectangular box in Fig.~\pref{f7}), tracking and isolating the wave motions on a ``single'' loop system delineated in the left hand panel from bottom to top (0 to 6). The timeseries of the Doppler velocity along that path (top panel), and the (nine timestep running average) trend-removed version (middle) is shown and we can see a pattern of upward propagating waves originating at {\em both} end points of the loop, in line with our earlier assumption. There is little evidence of waves traveling across the entire loop and no clear evidence of large-scale reflection near the points 1 and 5, far less the ``ends'' of the loop system at points 0 and 6. Comparing the velocity and line width timeseries we see little evidence of quasi-periodic changes in line width, but note the systematic variation of the latter across the loop \-- line widths growing by up to 5km/s in the central region where the two sets of outward propagating waves interact with one another \-- the low spatial resolution of CoMP makes it difficult to assess what is happening in and around the upper portions of the loop system although some of this line width growth must be associated with a density drop-off in the extended corona. Are the low-frequncy waves propagating through each other and interfering? Only detailed, high resolution, high cadence observations like those from ATST will be able to tell definitively.

Finally, a note on \alfvenic{} wave energy. CoMP resolves coronal \alfvenic{} wave amplitudes of $\sim$0.5km/s which, when compared to the $\sim$20km/s measured near the limb by \hinode{} and the \sdo{}, might, at first glance, seem like a large discrepancy hinting that there is really insufficient \alfvenic{} wave energy in the corona to energize or accelerate the plasma. \citet{2012ApJ...761..138M} used a blend of observational data and a simple forward model of \alfvenic{} wave propagation to resolve the apparent discrepancy and determine the \alfvenic{} wave energy content of the corona. Their results indicated that the enormous line-of-sight superposition within the coarse spatio-temporal sampling of CoMP hides the strong wave flux observed by \hinode{} and \sdo{} and leads to the large non-thermal line broadening observed. That exact scenario has been {\em assumed} in the past \citep[e.g.,][]{1990ApJ...348L..77H, 1994SSRv...70..373H, 1998ApJ...505..957C, 1998A&A...339..208B}. The strong correlation between the non-thermal line broadening and the wave amplitudes observed gives some confidence that the wave-related non-thermal line broadening is truly an indicator of a large wave energy reservoir for the heliosphere.

Unfortunately, on-disk determinations of these small-scale processes are difficult to make on a regular basis, requiring significant data analysis techniques as is clear from the discussion above. Fortunately, with observatories like HiC \citep[][]{2013Natur.493..501C,2013ApJ...770L...1T,2013ApJ...775L..32A,2013A&A...556A.104P,2013ApJ...771...21W} and {\em IRIS} we will have the spatial resolution to directly observe and characterize them and provide a more detailed look at the coupled small-scale processes filling the outer atmosphere. Unfortunately, there is no instrument that can reproduce CoMP-like observations of the corona on the solar disk and so portions of the energy spectrum contained in the velocity field of the coronal plasma (such as \alfvenic{} waves) will remain obscured from view on the global scale with only inference from the transverse motions visible (e.g., Fig.~\pref{f6}). Therefore an instrument of CoMP-like capability, and/or technology development along these lines, is absolutely necessary for the next generation of solar observatories if we are to constrain observationally driven models of the inner heliosphere that are required for space weather forecasting.

\section{Combining CoMP and SDO Observations in a Polar Plume}\label{comb} 
The previous sections have demonstrated the presence of ubiquitous phenomena in the quiescent solar atmosphere. Further, we have inferred that these phenomena work in tandem to produce the atmosphere observed and sustained it. However, we do not know how this one-two punch works and, because of the aforementioned limitations of on-disk coronal imaging spectroscopy, we are confined to the analysis of regions above the limb and (for now) the joint study of {\em SDO} and CoMP. The work presented by \citet{2013A&A...556A.124T} is the first study of this kind. In this following paragraphs we will perform a similar analysis to that of \citet{2013A&A...556A.124T} in a (apparently) simpler coronal hole environment to investigate wave and mass-flow properties as the previous measurements (Sect.~\pref{oscale}) have hinted at possible scenarios for the loading and acceleration of the fast solar wind from which a physically-motivated model of the fast wind {\em could} possibly be constructed.

The low spatial resolution of CoMP makes it less than ideal for studying the small intensity changes associated with the PCDs observed by \sdo/AIA (Fig.~\pref{f5}). Similarly AIA, by its design as a broadband imager, is not able to sense the subtle spectral changes of these \alfvenic{} waves running through the extended corona and so there is an area of discovery waiting for the next generation of spectral imaging experiments. For illustrative purposes however we consider the combination of CoMP and AIA in an open field region where we have the best chance to demonstrate the differences in mass and wave energy transport on the ``same'' structure without complications introduced in the analysis and interpretation by two-way flows and waves in the magnetically closed regions. Fortunately, there is sufficient Fe~XIII emission in a polar plume such that CoMP has enough signal to reliably measure Doppler shift and AIA can detect PCD propagation along the structure.

\begin{figure}
\begin{center}
\includegraphics[width=85mm]{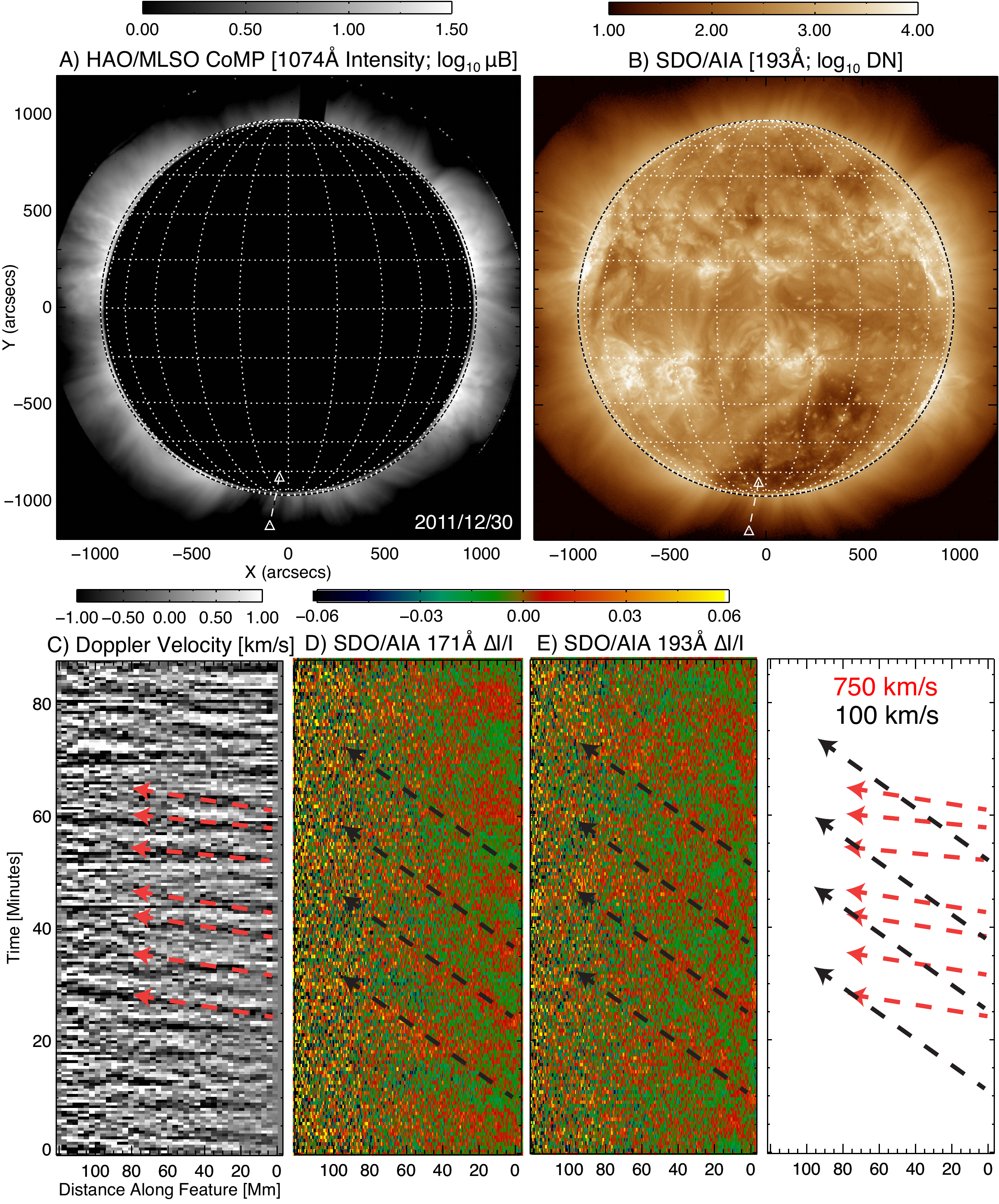}
\caption{CoMP and \sdo/AIA observations of a polar plume. Panels A and B show context image from the two instruments respectively showing a sample path on a polar plume rooted in the South polar coronal hole where we can assess some of the subtle dynamics in the plane of the sky and in the line of sight. Panels C through E show the space time plots on a polar plume from CoMP Doppler velocity, AIA 171\AA{} and 193\AA{} channels respectively. On each of these panels arrows are drawn to illustrate the propagating signals seen in each. These arrows are overlaid in the final panel for reference.}\label{f9} 
\end{center}
\end{figure}

The polar plume observations shown in Fig.~\pref{f9} demonstrate remarkable behavior. The (transverse) \alfvenic{} waves running outward in the CoMP Doppler shift observations have a strong $\sim$5-minute periodicity and a characteristic outward phase speed of 750km/s. The waves in the space-time plots show no appreciable acceleration in the CoMP field of view and very little inward wave propagation. The (longitudinal) PCDs observed in AIA intensity have apparent motions of $\sim$100km/s and periodicities of 15 minutes, similar to those discussed above and shown in Fig.~\pref{f5}. Further, we see that the apparent motion of PCDs in the 171 and 193\AA{} channels is the same, i.e. there is no clear temperature dependence in the propagation speeds, consistent with the analysis of \citet{2012ASPC..456...97T} and \citet{2012ASPC..455..361S}.

If we make the (reasonable) assumption that the PCDs represent outward propagating density enhancements {\em along\/} the magnetic field, then the transverse waves (running along the same structure and propagating at more than 5 times that speed) are experiencing a rapidly changing density structure in addition to the drop off of density along the plume. These outward propagating \alfvenic{} waves must experience reflection due to these density enhancements \-- a self-consistently occurring physical picture that can create persistent \alfvenic{} counter-propagation along the magnetic field. 

Observations such as these appear to show be ideal conditions for the onset of solar wind turbulence driven by low-frequency \alfvenic{} waves \citep[e.g.,][]{1991ApJ...372..719P, 2005ApJS..156..265C, 2005ApJ...632L..49S,2006ApJ...640L..75S,2007ApJS..171..520C,2007ApJS..171..520C,2010ApJ...708L.116V}. In such a picture, the waves persistently propagate outward along the inhomogenous structure, some portion of the wave power is reflected by each ``plug'' of material, the counter-probating waves interact driving a cascade to higher frequencies and smaller scales \citep[see, e.g.,][]{2013PhRvL.110i1102K} which create a wave pressure needed to push the material faster. The interested reader is referred to a more detailed combined study of CoMP and AIA polar plume observations (McIntosh, Bethge \& Tian 2014, in preparation) and the effect of the propagating inhomogeneity on the wave field (McIntosh et al. 2014, in preparation) and solar wind that is based on the previously published model of \citet{1998JGR...10323677O}.

\section{Summary}
While it is unclear from direct observation if the coronal magnetic field is intrinsically turbulent, the processes which govern its shape and evolution in the solar interior and lower solar atmosphere certainly are. The two, possibly three, scales of magneto-convective granulation constantly force the quiescent portions of the outer atmosphere on timescales of minutes, days, (and weeks/months) such that the outer atmosphere that is never in a stationary state \-- only one that is always trying to relax. 

In the process of the relentless relaxation across scales small discontinuities in the magnetic field occur and a considerable amount of energy from that stressed field {\em must\/} be released into the plasma, heating and accelerating it as is the picture of ``nanoflares'' proposed by Parker. Indeed, \citet{2011Sci...331...55D} had already deduced that Type-II spicules (and PCDs), based on their ubiquity and estimates of their energy content, are the manifestation of Parker's hypothesis, but in the chromosphere close to where the hydrodynamic and magnetic pressures balance. 

The other portion of the energy budget, the second stage if you like, comes from the convective velocity spectrum and the generation of torsional/\alfvenic{} motions that are clearly ubiquitous in the outer atmosphere, riding outward along {\em all} chromospheric and coronal structures. We believe it is the potential for counter-propagating \alfvenic{} waves through a structure riddled with inhomogeneity that {\em should\/} give rise to an isotropic heater for the coronal plasma \citep[e.g.,][]{2012RSPTA.370.3217P} and an accelerating base for the solar wind. 

We have high hopes that the next generation of photospheric, chromospheric and coronal observations (like those planned for {\em IRIS}, ATST and the planned {\em Solar-C} mission) will open new frontiers in the reconciliation of mass and wave transport at the constantly evolving magneto-thermodynamic base of the heliospheric system. We similarly hope that some of the speculation presented above provides a spur for future investigations of how the magnetic scales of the outer atmosphere cascade downward to power the energetic processes which eventually populate the heliosphere.

\begin{acknowledgements}
SMC would like to thank Bart De Pontieu for frequently discussing the issues outlined in this manuscript. We apologize for not thoroughly reviewing the vast literature on the topics covered---hopefully the references included provide an adequate cross-section of the excellent work done in this area. The material presented was supported by the National Aeronautics and Space Administration under grant NNX08AU30G issued by the Living with a Star Targeted Research \& Technology Program. In addition, part of the work is supported by NASA grants NNX08AU30G, NNX08AL23G, NNM07AA01C \-- {\em Hinode}, NNG09FA40C \-- {\em IRIS} and ATM-0925177 from the National Science Foundation. IDM acknowledges support of a Royal Society University Research Fellowship.  The research leading to these results has also received funding from the European Commissions Seventh Framework Programme (FP7/2007-2013) under the grant agreements SOLSPANET (project No. 269299, www.solspanet.eu/solspanet). The National Center for Atmospheric Research is sponsored by the National Science Foundation. {\em Hinode\/} SOT/SP Inversions were conducted at NCAR under the framework of the Community Spectro-polarimetric Analysis Center (CSAC; \url{http://www.csac.hao.ucar.edu/}).
\end{acknowledgements}


\end{document}